\def\Slash#1{\ooalign{\hfil/\hfil\crcr$#1$}}
\documentclass{PoS}

\title{Hadronic description for Omega baryon photoproduction}

\ShortTitle{Omega baryon photoproduction}

\author{\speaker{Hui-Young Ryu} \\
        Korea Institute of Science and Technology Information (KISTI) 245,
 Daejeon 305-806, Korea\\
        E-mail: \email{hyryu@kisti.re.kr} }

\author{Atsushi Hosaka\\
        Research Center for Nuclear Physics, Osaka University,
  Ibaraki 567-0047, Japan \\
        E-mail: \email{hosaka@rcnp.osaka-u.ac.jp}}

\author{H. Haberzettl\\
        Institute for Nuclear Studies and Department of Physics,
  The George Washington University, Washington, DC 20052, USA\\
        E-mail: \email{helmut@gwu.edu}}

\author{Hyun-Chul Kim\\
        Department of Physics, Inha University, Incheon 402--751, 
Republic of Korea\\
        School of Physics, Korea Institute for Advanced Study,
Seoul 130--722, Republic of Korea\\
        E-mail: \email{hchkim@inha.ac.kr}}

\author{K. Nakayama\\
        Department of Physics and Astronomy, The University of
  Georgia, Athens, GA 30602, USA\\
Institue f\"ur Kernphysick and Center for Hadron Physics,
Forschungszentrum J\"ulich, 52424 J\"ulich, Germany \\
        E-mail: \email{nakayama@uga.edu}}

\author{Yongseok Oh\\
        Department of Physics, Kyungpook National University,
  Daegu 702-701, Korea\\
Asia Pacific Center for Theoretical Physics, Pohang,
  Geyoengbuk 790-784, Korea \\
        E-mail: \email{yohphy@knu.ac.kr}}

\abstract{In the present work, we investigate subsequential production of three kaons
and $\Omega^-$ baryon based on an effective Lagrangian
approach. We only consider the intermediate states with the light mass
baryon to suggest the minimum of the total cross section.
Coupling constants for verteces of meson-octet baryons are
fixed from the empirical data and/or quark models together with SU(3)
symmetry considerations and these for meson-decouplet are predicted
not only quark model but also Chiral-quark soliton model calculation.
Gauge invariance of the resulting amplitude
is maintained by introducing the contact currents by extending the
gauge-invariant approach of Haberzettl for one-meson photoproduction
to two-meson photoproduction.}

\FullConference{XV International Conference on Hadron Spectroscopy-Hadron 2013\\
		4-8 November 2013\\
		Nara, Japan }

\begin{document}
\section{Introduction}
$\Omega^-$ baryon $(sss)$ was predicted by the quark model in 1962 and
its existance was proved experimentally in 1964 \cite{Barnes:1964}. 
However, since the
laste 1980's, few significant progress has been made in Oemga
spectroscopy because of the closing of the then existing kaon
factories. 
In 2006, the spin property of the omega baryon was shown \cite{Aubert:2006dc}.
Recently, the CLAS Collaboration at the Thomas Jefferson
National Accelerator Facility (JLab) initiated a cascade physics
program; the Collaboration has established, in particular, the
feasibility to do omega baryon spectroscopy via photoproduction
reactions such as $\gamma p \to K^+ K^+ K^0 \Omega^-$ \cite{Strakovsky_proposal}.
A dedicated experiment for this reaction is currently underway.
In the present work, we would like to suggest the extimation of the
toall cross section of the omega baryon production.
\section{Formalism}
\begin{figure}[ht]
    \begin{minipage}{.4\textwidth}
       \includegraphics[width=6cm]{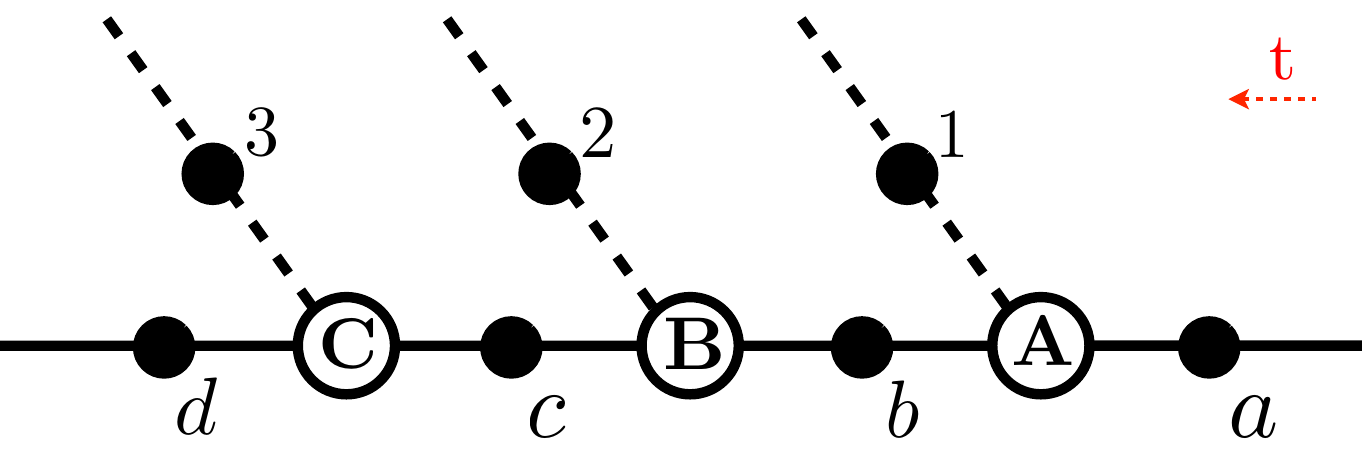}
    \end{minipage}  
    \begin{minipage}[]{.6\textwidth}
  \caption{Sequential hadronic photoproduction mechanism of three
    kaons off an initial baryon.} \label{fig:1}
    \end{minipage}
\end{figure}
The meson-baryon diagram we considered In the present work is shown in Figure
\ref{fig:1}.  
Sequential hadronic photoproduction mechanism of three
    kaons off an initial baryon $B_a \to M_1 + M_2 +M_3 +B_d$. Indices
    $x=a,b,c,d$ label the baryons $B_z$ depicted as solid lines and
    $y=1,2,3$ label the $K$ mesons $M_y$ shown as dashed lines. The
    three meson-baryon-baryon vertices $F_Z$ are labeled by $Z=A,B,C$;
    the vertex $F_A$, for example, describes the transition $B_a \to
    M_1 + M_b$.
Acorrding to the position of the nuetural kaon, there can
be three type of diagram such that $(M_1, M_2, M_3)$ can be
$(K^+,K^+,K^0)$, $(K^+,K^0,K^+)$ and $(K^0,K^+,K^+)$. 
Denoting the four single-baryon currents by $\Gamma_x^\mu$ (where
$x=x,b,c,d$), the three single-meson currents by $J_y^\mu$ (where
$y=1,2,3$), and the three contact-type interaction cuttrents by
$M_Z^\mu$ (where $Z=A,B,C$), the total three-meson production current
resulting from this procedure has ten topologically distinct contributions,
\begin{eqnarray}
  \label{eq:1}
  M^\mu &=& \underbrace{F_C t_c F_B t_b F_A t_a \Gamma^\mu_a + F_C t_c F_B t_b
  \Gamma_b^\mu t_b F_A + F_C t_c \Gamma_c^\mu t_c F_B t_b F_A +
  \Gamma_d^\mu t_d F_C t_c F_B t_b F_A }_{\textrm{baryon
    currents}} \nonumber \\*[.2cm]
&{}& + \underbrace{F_C t_c F_B t_b J_1^{\mu} \Delta_1 F_A + F_C t_c J_2^\mu
\Delta_2 F_B t_b F_A + J_3^\mu \Delta_3 F_C t_c F_B t_b F_A
}_{\textrm{meson currents}} \nonumber \\*[.2cm]
&{}& + \underbrace{F_C t_c F_B t_b M_A^\mu + F_C t_c M_B^\mu t_b F_A + M_C^\mu t_c
F_B t_b F_A }_{\textrm{interaction currents} } 
\end{eqnarray}
where $t_x ~(x=a,b,c,d)$ and $\Delta_x ~(y=1,2,3)$ are the baryon and
meson propagators, respectively.

In addition to this meson-baryon diagram, the photon
can couple to 8 positions except the neutral kaon line. 
Therefore we consider 24 diagrams. To suggest the minimum of the total cross
section, we only consider the lighest hyperon states.

\newpage
As a one example, let us consider one baryon current in the case of
$(M_1, M_2, M_3) = (K^+,K^+,K^0)$. The vertex functions in this case
are given by
\begin{eqnarray}
  \label{eq:2}
F_{\Xi} &=& g_{\Xi} p_{3\lambda} f_{\Xi} (p_3^2;p_4^2,q_2^2), \\*[.2cm]
t_{\Xi}  &=& \frac{\Slash{q}_2 + m_{\Xi}}{q_2^2 -m_{\xi}^2} ,\\*[.2cm]
F_{\Lambda}  &=& g_{\Lambda} \gamma_5 \Slash{p}_2 f_{\Lambda} (p_2^2;q_2^2,q_1^2) , \\*[.2cm]
t_{\Lambda}  &=& \frac{\Slash{q}_2 + m_{\Xi}}{q_1^2 -m_{\Lambda}^2} ,\\*[.2cm]
F_p  &=&  g_{p} \gamma_5 \Slash{p}_1 f_{p}
(p_1^2;q_1^2,q_3^2) , \\*[.2cm]
t_p  &=&  \frac{\Slash{q}_3 + m_{p}}{q_3^2 -m_{p}^2}   , \\*[.2cm]
\Gamma_p  &=& \bigg[ I + \frac{\kappa_p}{2m_p} \Slash{k}_1 \bigg]
\Slash{\epsilon}_\gamma
\end{eqnarray}
where $k_1$ and $k_2$ are momentum of incoming photon and proton, respectively;
$p_1$, $p_2$, $p_3$ and $p_4$ are the outgoing three kaons and omega
baryon, respectively; other momentum $q_i$ and $\epsilon_\gamma$ are
momentum of the intermediate hyperon and the polarization vector of
the photon.
In above case, the photon couples to the incoming proton.

\section{Numerical result}
\begin{figure}[ht]
    \begin{center}
       \includegraphics[width=10cm]{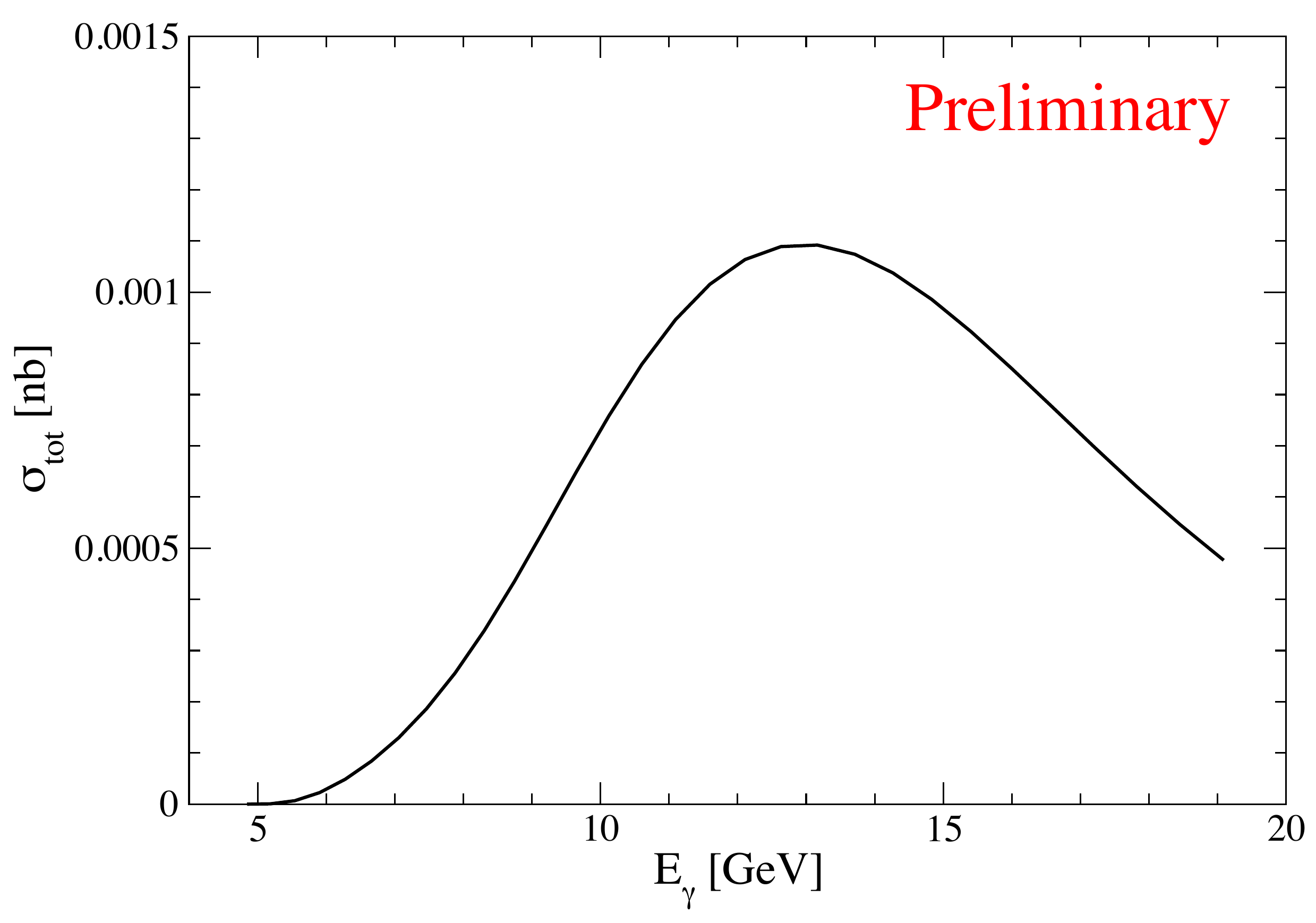}
       \vspace{-10pt}
  \caption{Total cross section as a function of the photon energy. } \label{fig:2}
    \end{center}
\end{figure}
In Figure \ref{fig:2}, we show the total cross section up to 20
GeV to see when the cross section starts to
decrease even though effective Lagrangian approach is ambiguous at
such a high energy region. Parameters in this work are taken
from Ref. \cite{Nakayama:2006ty} for baryons octet
and Ref. \cite{Yang:2013qza} and \cite{Kim_private} for baryon decuplet.

\section{Summary and outlook}
In the present talk, we reviewed a recent study of $\Omega$ baryon photoproduction off the
nucleon target, i.e., $\gamma p \to K^+ K^+ K^0 \Omega^-$. Since there
is no data of the cross section for $\Omega^-$ baryon, we suggest the
minimum only considering the lightest mass hyperon intermediate
states.
In the future, we would like to include the role of high-spin hyperon
resonances cotribution in this scattering process.


\begin{thebibliography}{99}

\bibitem{Barnes:1964} 
  V.~E.~Barnes, P.~L.~Connolly, D.~J.~Crennell, B.~B.~Culwick, W.~C.~Delaney, W.~B.~Fowler, P.~E.~Hagerty and E.~L.~Hart {\it et al.},
  Phys.\ Rev.\ Lett.\  {\bf 12}, 204 (1964).


\bibitem{Aubert:2006dc} 
  B.~Aubert {\it et al.}  [BaBar Collaboration],
  Phys.\ Rev.\ Lett.\  {\bf 97}, 112001 (2006)
  [hep-ex/0606039].



\bibitem{Strakovsky_proposal}
\textit{Photoproduction of the Very Strangest Baryons on a Proton Target in
CLAS12}
Spokenperson : I.I. Strakovsky, proposal (version: 3 June 2013)


\bibitem{Nakayama:2006ty} 
  K.~Nakayama, Y.~Oh and H.~Haberzettl,
  Phys.\ Rev.\ C {\bf 74}, 035205 (2006)
  [hep-ph/0605169].


\bibitem{Yang:2013qza} 
  G.~-S.~Yang and H.~-C.~Kim,
  Few Body Syst.\  {\bf 54}, 325 (2013).


\bibitem{Kim_private} 
H. C. Kim (private communication)

\end{thebibliography}
\end{document}